\documentclass[10pt,letter,final]{IEEEtran}
\usepackage{cite}
\usepackage[final]{graphicx}
\usepackage{amsthm}

\usepackage{amsmath,amsfonts}

\def\bH{\mathbf H}
\def\tj{\tilde j}
\def\tk{\tilde k}
\def\cD{{\mathcal D}}
\def\cDin{{\mathcal D}^{\text{(in)}}}
\def\cJ{{\mathcal J}}
\def\cT{{\mathcal T}}
\def\hj{{\hat j}}
\def\hk{{\hat k}}

\def\cI{{\mathcal I}}
\def\cK{{\mathcal K}}

\def\bard#1#2{\bar d_{#1,#2}}
\def\setZ{{\mathbb Z}}
\def\setR{{\mathbb R}}

\def\paren#1{\left(#1\right)}
\DeclareMathOperator{\bydef}{:\!=}
\def\part#1{\medskip\noindent\textsc{#1}}
\newtheorem{theorem}{Theorem}
\def\figref#1{Fig.~\ref{#1}}
\def\secref#1{Sec.~\ref{#1}}
\def\thmref#1{Sec.~\ref{#1}}
\def\QED{{\ensuremath \Box}}
\usepackage[notcite,notref]{showkeys}

\makeatletter
\def\subsubsection{
     \@startsection {subsubsection}{3}{\z@ }%
     {-1.25ex\@plus -1ex \@minus -.2ex}%
     {0.6ex \@plus .4ex \@minus .1ex}%
     {\itshape}%
} \makeatother

\begin{document}

\title{Real Interference Alignment and Degrees of Freedom Region of
Wireless X Networks}

\author{Zhengdao~Wang\\
\small
Department of Electrical and Computer Engineering, Iowa State
University, Ames, IA 50011 USA\\
e-mail:  zhengdao@iastate.edu}

\maketitle
\begin{abstract}
We consider a single hop wireless X network with $K$ transmitters and $J$
receivers, all with single antenna. Each transmitter conveys for each receiver
an independent message. The channel is assumed to have constant coefficients.
We develop interference alignment scheme for this setup and derived several
achievable degrees of freedom regions. We show that in some cases, the derived
region meets a previous outer bound and are hence the DoF region. For our
achievability schemes, we divide each message into streams and use real
interference alignment on the streams. Several previous results on the DoF
region and total DoF for various special cases can be recovered from our
result.

\end{abstract}

\linespread{.99}\normalsize

\section{Introduction} The wireless X network \cite{caja09}, models a
single-hop wireless network such that each transmitter conveys an independent
message for each receiver. All transmitter and all receivers have single
antenna. Multiple antenna extensions have been considered \cite{jash08}. The X
network model includes the broadcast channels, multiple access channels, and
the interference channels as special cases. It is therefore useful to quantify
the capacity limits of $X$ networks. However, this is a difficult problem
because even the capacity region for the broadcast channel, which is a special
case of the X network, has not been characterized in full generality (e.g.,
discrete memoryless broadcast channel).

Simple single-letter type characterizations of capacity regions for many of
other multi-user information-theoretic problems have also eluded us. A recent
line of attack focuses on Gaussian networks in the asymptotic regime where the
signal to noise ratio (SNR) goes to infinity. The communication rates are
normalized by $\log(\text{SNR})$ to yield a quantity defined as the degrees of
freedom (DoF), or multiplexing gain \cite{zhts02}. The shape of the capacity
region normalized by $\log(\text{SNR})$ as SNR goes to infinity is defined as
the DoF region, e.g., \cite{jash08}. The total DoF and in some cases the DoF
region for several channels have been characterized recently. One important
technique for proving the achievability results is the \emph{interference
alignment}, which seeks to align the dimensions of interference signals so
that more dimensions are available in the subspace unaffected by interference.

There are several interference alignment techniques, among which are the
vector interference alignment based on beamforming and zero-forcing, e.g.,
\cite{mamk06c,caja08}, and the real interference alignment
\cite{brpt10,sjvj08c,etor09,mgmk09}. There seems to be intimate connections
between the two methods.

For the DoF problem of wireless X network, several results are available. An
outer bound for multiple-input multiple-output (MIMO) X network has been
derived in \cite{caja09}, which also developed schemes for achieving the
maximum total DoF for single antenna X network. For constant single-antenna
channels, a real interference alignment scheme has been used in \cite{mgmk09}
to establish the maximum total DoF. For MIMO X networks, outer bounds and
achievability schemes have been developed in \cite{jash08} for the $2\times 2$
MIMO X network. The DoF region for an $M\times 2$ X network with $N_1$ and
$N_2$ antennas at the two receivers is available as a special case of the
result in \cite{yikw11i}. Antenna splitting argument has been used in
\cite{caja09} to establish a lower bound on the total DoF of MIMO X network.

In this paper, we focus on the single-antenna wireless X networks, and derive
several achievability schemes based on real interference alignment. The
achieved DoF regions are shown to be tight when the number of receivers is
two. Several previous results (or their constant channel counter parts) can be
recovered as special cases.

\section{System Model}\label{sec.model}

Notation: Throughout the paper, $J$ and $K$ will be integers and $\cJ=\{1,
\ldots, J\}$, $\cK=\{1, \ldots, K\}$. We use $k$, $\tk$, $\hk$ as transmitter
indices, and $j$, $\tj$, $\hj$ as receiver indices. The set of integers and
real numbers are denoted as $\setZ$ and $\setR$, respectively. We use
$[d_{j,k}]$ to denote a matrix with element $d_{j,k}$ in the $(j,k)$th
position, and use $[d_{j,k}]_{j=1,k=1}^{J,K}$ to make the size of the matrix
explicit. Letter $l$ will be reserved for the index of streams (parts of a
message). Throughout the paper, a.e.~means almost everywhere in the Lebesgue
sense for the channel matrix. \hfill \QED

Consider a single-antenna wireless X network with $K$ transmitters and $J$
receivers. For each pair $(j,k)\in\cJ\times \cK$, transmitter $k$ conveys a
message $m_{j,k}$ for receiver $j$. The channel from transmitter $k$ to
receiver $j$ is denoted as $h_{j,k}$. The whole set of channel coefficients is
denoted as a matrix
\begin{equation}
\bH\bydef [h_{j,k}]_{j=1,k=1}^{J,K}.
\end{equation}
All the quantities are real in this paper. So $\bH\in \setR^{J\times K}$. The
channel is assumed constant (non-fading) throughout the whole transmission.
Each transmitter $k$ transmits a symbol $x_{k,t}$ in time slot
$t\in\setZ$. Each transmitter has an average power constraint $P$ so that for
any transmission that spans $N\in\setZ$ symbols, the transmitted symbols
satisfy
\begin{equation}
  \sum_{t=1}^N \frac 1N |x_{k,t}|^2 \le P, \quad \forall 1\le k\le K.
\end{equation}
The received signal at receiver $j$ at time $t$ can be written as
\begin{equation}
  y_{j,t}= \sum_{k\in \cK} h_{j,k} x_{k,t} + \nu_{j,t}, \quad \forall j\in \cJ
\end{equation}
where $\{\nu_{j,t}|j\in \cJ\}$ is the set of additive noises, assumed to be
independent and identically distributed according to zero mean Gaussian
distribution with unit variance. So $P$ is the per-message SNR.

A code of length $N$ and message sizes $[M_{j,k}]$ consists of
\begin{enumerate}
\item the encoders $\{f_k|k\in\cK\}$, where $f_k$ is a mapping from
  the set of messages to be conveyed by transmitter $k$,
  $[1,M_{1,k}]\times,\ldots, \times [1,M_{J,k}]$,
  to the set of transmitted symbols (codewords) in $\setR^N$. All codewords satisfy the
  power constraint.
\item the decoders $\{g_{j,k}|j\in \cJ, k\in\cK\}$, where $g_{j,k}$ is a
mapping from the set $\setR^N$ of received symbols at receiver $j$ to the set
of messages $[1,M_{j,k}]$ intended for receiver $j$ from transmitter $k$.
\end{enumerate}
The rate of message $m_{j,k}$ is defined to be
\begin{equation}\label{eq.rate1}
R_{j,k}=\frac 1N \log_2(M_{j,k}).
\end{equation}
Let $[W_{j,k}]$ denote a set of messages such that $W_{j,k}$ is independently
and uniformly chosen from $[1, M_{j,k}]$. The probability of error $P_e$ of
the code is defined as
\begin{align*}
  \text{Pr}\left[g_{j,k}\left(\sum_{k\in\cK} h_{j,k} f_k
  ([W_{j,k}]_{j=1}^J) + [v_{j,t}]_{t=1}^N \right) \ne W_{j,k}\right.\\
  \text{for some } (j,k)\in \cJ\times\cK\biggr].
\end{align*}
The code we have thus defined will be denoted as a
\begin{equation} \label{eq.code}
(P, N, [M_{j,k}], [f_k], [g_{j,k}])
\end{equation}
code.

The degree of freedom (DoF) region for the system is the closure of a set of
points $[d_{j,k}]\in\setR^{J\times K}$ such that for any $\epsilon>0$, there
is a sequence of codes $(P^{(i)}, N^{(i)}, [M_{j,k}^{(i)}], [f_k^{(i)}],
[g_{j,k}^{(i)}])$ indexed by $i$ such that as $i\to\infty$, the power
$P^{(i)}\to\infty$, and
\begin{equation}\label{eq.rate2}
  \lim_{i\to\infty}\frac{R_{j,k}^{(i)}}{0.5\log P^{(i)}}=d_{j,k}, \quad
  \forall j\in\cJ, \forall k\in \cK,
\end{equation}
and such that for all $i$, the probability of error is less than $\epsilon$.

\section{Achievable Degrees of Freedom Region} \label{sec.dof}

\subsection{Statement of result}

\begin{theorem}[An achievable DoF region] \label{th.main}
For a $K$-transmitter $J$-receiver constant-coefficient single-antenna
wireless X network $\bH\in \setR^{J\times K}$, the DoF region $\cD$ satisfies
$\cD\supset \cDin$ a.e., where $\cDin$ is a set of matrices $[ d_{j,k}
]_{j=1,k=1}^{J,K} $ such that
\begin{enumerate}
\item all entries of it are non-negative;
\item $\forall 1\le j\le J$, the following inequality holds:
\begin{equation} \label{eq.main}
  \sum_{\tk=1}^K d_{j,\tk} + \sum_{\tj\in \cJ,\tj\ne j} \max_{k} d_{\tj, k}
  \le 1.
\end{equation}
\end{enumerate}
\end{theorem}

\subsection{Main ideas} Our achievability proof uses the following ideas:
\begin{enumerate}
\item We use \emph{real interference alignment}, a technique that has been
initiated in \cite{brpt10}, and further extended for the interference problems
in \cite{sjvj08c,etor09,mgmk09}.
\item We split each message into \emph{streams}, where all streams have the
same have the same DoF. This allows us to design achievability schemes for
unequal DoFs. This idea has been used in e.g., \cite{krwy11i}.
\item The interference alignment at the receivers is stream-based. Several
streams from different transmitters are aligned. Streams from the same
transmitter are never aligned. Otherwise decodability of the aligned messages
at other receivers will be compromised.
\item We use a construction that involves ``dimension padding'' to guarantee
that all streams have the same DoF.
\end{enumerate}

\subsection{The proof}
\begin{figure*}
\centering
\includegraphics[width=.7\textwidth]{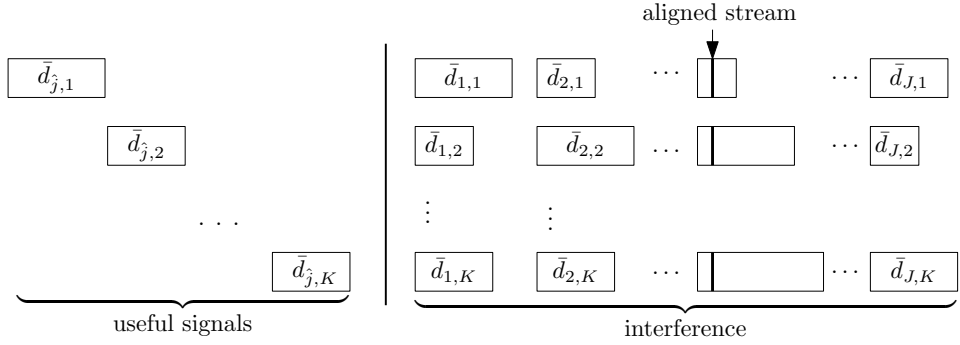}
\caption{Interference alignment at receiver $\hj$}
\label{fig.align}
\end{figure*}

We prove that for any $[d_{j,k}]\in \cDin$, $[d_{j,k}]$ is achievable. We
assume that all the elements of $[d_{j,k}]$ are rational numbers. Otherwise,
if some elements are irrational, the proof here can be used to establish
achievability of a point that is arbitrarily close to $[d_{j,k}]$.

Under the rational assumption, we can find an integer $\kappa$ such that for
all $j\in \cJ$ and all $k\in \cK$, $\bar d_{j,k}\bydef \kappa d_{j,k}$ is a
non-negative integer.

\part{Encoding}: For each $(j,k)\in \cJ\times \cK$, the message $m_{j,k}$ is
divided into $\bard jk$ parts as $\{m_{j,k,l}, l=1, \ldots, \bard jk\}$. Each
part is called a \emph{stream}. The signal emitted by transmitter $k$ is in
the following form
\begin{equation}
x_k = \sum_{j\in \cJ} x_{j,k}
  = \sum_{j\in \cJ} \sum_{l=1}^{\bard jk} \alpha_{j,k,l} x_{j,k,l}
\end{equation}
where $x_{j,k,l}$ carries the symbols of stream $l$ of the message from
transmitter $k$ to receiver $j$, and $\{\alpha_{j,k,l}\}$ are design
parameters that can be chosen randomly, independently, and uniformly from the
interval $[\frac 12, 1]$. The symbol $x_{j,k,l}$ is generated using elements
(called \emph{directions} \cite{mgmk09}) specified in a set $\cT_{j,k,l}$ (to
be specified later) as follows:
\begin{equation}\label{eq.stream}
  x_{j,k,l} = \sum_{\delta_b\in \cT_{j,k,l}} \delta_b u_{j,k,l,b}
\end{equation}
where $u_{j,k,l,b}\in \{\lambda q| q\in \setZ, -Q\le q \le Q\}$, and
$Q$ and $\lambda$ are parameters to be specified appropriately later to
satisfy the rate and power requirements. In the summation in
\eqref{eq.stream}, we have assumed that the directions in $\cT_{j,k,l}$ have
been indexed from $1$, and $b$ is the index of the direction of $\delta_b$. The
exact indexing scheme is of no importance.

\part{Stream alignment}: Consider an arbitrary receiver $\hj$. The signal
dimensions situation is shown in \figref{fig.align}. The useful signals have
DoF $\sum_{k\in \cK} \bard {\hj}k$. The interferences coming from different
transmitters are shown on the right. The streams intended for the same
receiver $j\ne \hj$ are aligned together at receiver $\hj$.

\part{Dimension padding}: To facilitate the construction of the transmission
directions, we introduce an idea that we term \emph{dimension padding}.
Specifically, we notice that in the interference part in \figref{fig.align},
the messages intended for the same receiver $j\ne \hj$ in general do not have
the same number of streams. To make sure that such disparity does not lead to
difference in the achieved DoF for these messages, we introduce some
fictitious streams so that with these additional streams the constructed
transmission symbols for all \emph{actual} streams use the same number of
directions. These fictitious streams only aid in the construction of the
transmission directions. No symbols are transmitted for these streams,
otherwise the useful signal space dimension will become smaller (the
interference space dimension remains unchanged though).

More specifically, we assume all messages $\{m_{j,k}|k\in \cK\}$ intended for
receiver $j$ has the same number $s_j$ of streams, where
\begin{equation}
  s_j = \max_k {\bard jk}.
\end{equation}
For transmitter $k$, the first $\bard jk$ of these $s_j$ streams are actual
transmitted streams. The remaining ones (if any) are virtual streams, whose
transmitted symbols are all set to zero [c.f.~\eqref{eq.stream}]:
\begin{equation}
u_{j,k,l,b} = 0, \quad \forall l \in [\bard jk +1, s_j].
\end{equation}
We assume that $\alpha_{j,k,l}$ is assigned for a virtual stream in the same
way as for an actual stream.

\part{Transmit directions}: We design the directions $\cT_{j,k,l}$ used by
stream $m_{j,k,l}$ to contain and only contain directions of the following
form:
\begin{equation}\label{eq.T}
T=\prod_{\hj\in \cJ,\hj\ne j} \prod_{\hk\in\cK}
  \paren{h_{\hj,\hk}\alpha_{j,\hk,l}}^{p_{\hj,\hk,j,k,l}}
\end{equation}
where
\begin{equation}
0\le p_{\hj,\hk,j,k,l} \le n-1,
\end{equation}
$\forall \hj\in \cJ, \hj\ne j$, $\forall \hk\in \cK$. It can be seen that
there are totally $n^{K(J-1)}$ directions in $\cT_{j,k,l}$ for all $(j,k,l)$.
The reason for doing dimension padding can be seen more clearly now as it
leads to the same number of directions to be used by all streams. This will
guarantee that each stream corresponds to the same DoF in the final result.

\part{Alignment verification}: The proposed design above guarantees that the
interferences created by messages intended for the same receiver are aligned
as desired at all receivers. To see this, define $\cT_{\hj,j,l}$ to contain
directions described by \eqref{eq.T} but with
\begin{equation}
0\le p_{\hj,\hk,j, k, l} \le n,
\end{equation}
for all $(\hj,\hk,j,k,i)$. According to \eqref{eq.stream}, a symbol from
stream $(j,k,l)$ is transmitted in a direction of the form $\alpha_{j,k,l} T$
where $T$ is as in \eqref{eq.T}. This symbol will arrive at receiver $\hj$ in
the direction of $\paren{h_{\hj,k}\alpha_{j,k,l}}T$, which is in
$\cT_{\hj,j,l}$ because the power for $\paren{h_{\hj,k}\alpha_{j,k,l}}$ will
be simply increased by one after the symbol goes through the channel. Note
that not all directions in $\cT_{\hj,j,l}$ will be occupied by interference so
the effective number of interference dimensions is smaller than the number of
elements in $\cT_{\hj,j,l}$. However, this does not affect the calculation of
the achievable DoF.

\part{Decodability}: The useful signals at receiver $\hj$ will be generated by
directions in $\cT'_{\hj}$, where
\begin{equation}
  \cT'_\hj = \bigcup_{k\in\cK} \paren{h_{\hj,k}\alpha_{\hj, k, l}}
  \cT_{\hj,k,l}\;.
\end{equation}
Since none of the $\cT_{\hj,k,l}$ contains a generator
$\paren{h_{\hj,k}\alpha_{\hj, k, l}}$ [recall the condition $\hj\ne j$ in
\eqref{eq.T}], and for different $k$, $\paren{h_{\hj,k}\alpha_{\hj, k, l}}$
are different, we conclude that $\cT'_\hj$ is rationally independent of
$\bigcup_{j,l} \cT_{\hj,j,l}$. Therefore, all the useful signals are decodable
in the noiseless case a.e..

The total rational dimensions $D_\hj$ of both the useful signals and the
interference at any receiver $\hj$ is
\begin{equation}
  D_\hj \le \sum_{\tk=1}^K \bar d_{\hj,\tk} n^{K(J-1)} +
  \sum_{j\in \cJ,j\ne \hj} \max_{k} \bar d_{j, k} (n+1)^{K(J-1)}. \notag
\end{equation}
We define
\begin{equation}
S= \max_{\hj\in \cJ}
  \paren{
  \sum_{\tk=1}^K \bar d_{\hj,\tk} + \sum_{j\in \cJ,j\ne \hj} \max_{k} \bar d_{j, k}},
\end{equation}
which is an upper bound on the total number of useful signal streams and
interference streams (multiple aligned streams are counted as one), maximized
over all receivers. For any DoF point in $\cDin$ that satisfies
\eqref{eq.main}, we have $S\le \kappa$. As a result, we have
\begin{equation}
  D_\hj \le S (n+1)^{K(J-1)} \le \kappa (n+1)^{K(J-1)}
\end{equation}

With reference to the constellation symbols in \eqref{eq.stream}, if we choose
\begin{equation}
\lambda = P^{\frac 12}/{Q}
\end{equation}
then we can guarantee that the power constraint is satisfied. In addition, if
for any $\epsilon \in (0, 1)$ we choose as in e.g., \cite{mgmk09},
\begin{equation}
Q=P^{\frac{1-\epsilon}{2(m+\epsilon)}},
\end{equation}
where $m$ is an integer, then we can guarantee that the DoF per stream is
$\frac{1-\epsilon}{m+\epsilon}$. Choosing $m=\kappa (n+1)^{K(J-1)}$, the hard
decoding error probability for the constellation symbols decreases to zero as
$P\to\infty$ due to the Khintchine-Groshev type Theorems, see the discussion
in e.g., \cite{mgmk09,etor09}, and the DoF of the message $m_{j,k}$ can be
arbitrarily close to
\begin{equation}
  \lim_{n\to\infty}\frac{\bard jk n^{K(J-1)}}{\kappa (n+1)^{K(J-1)}} =
  \frac{\bard jk}{\kappa}=d_{j,k},
\end{equation}
for all $j\in \cJ$ and $k\in \cK$. This completes the proof. \hfill \QED

\section{Extensions} \label{sec.ext}

The alignment scheme presented in \secref{sec.dof} is only one possible
alignment schemes within the class of real alignment. We have aligned the
messages intended for the same receiver. However this is not always optimal
and not necessary either. We propose some extensions of the alignment scheme
that can yield potentially larger achievable DoF regions.

\subsection{Permuted alignment} To see the insufficiency of the alignment
scheme in \secref{sec.dof}, consider a $3\times 3$ X network. If we set all
messages $m_{j,k}$ to have rate zero whenever $j\ne k$, then it becomes a
3-user interference channel. It is known \cite{caja08} that per user DoF 1/2
is achievable. Therefore, the following DoF point is within the DoF region of
the $3\times 3$ X network:
\begin{equation}\label{eq.half}
  [d_{j,k}]^T=\begin{bmatrix}
  \frac 12 & 0 & 0 \\
  0& \frac 12 & 0 \\
  0& 0& \frac 12 \\
  \end{bmatrix}.
\end{equation}
However, it can be seen that this point cannot be achieved using the scheme in
\secref{sec.dof}. To achieve this point, we can arrange the individual DoFs in
each row so that it looks as follows (c.f.~\figref{fig.align}):
\begin{equation}
\begin{bmatrix}
  d_{1,1} & d_{2,1} & d_{3,1} \\
  d_{2,2} & d_{1,2} & d_{3,2} \\
  d_{3,3} & d_{1,3} & d_{3,3}
\end{bmatrix}=
\begin{bmatrix}
  \frac 12 & 0 & 0 \\
  \frac 12 & 0 & 0 \\
  \frac 12 & 0 & 0
\end{bmatrix}.
\end{equation}
Note the matrix has been shown in its transposed form to agree with the
\figref{fig.align}. The permutations applied to different rows can be
different. To see that this point is achievable, we can check e.g., the
situation at receiver 1 as depicted (for illustration only) in the following
\begin{equation}
\begin{bmatrix}
  \frac 12 & - & - \\
   -  & 0 &  - \\
   -  & - & 0
\end{bmatrix}
\biggr|
\begin{bmatrix}
  - & 0 & 0 \\
  \frac 12 & - & 0 \\
  \frac 12 & - & 0
\end{bmatrix}
\end{equation}
where the left part is the signal dimensions, and the right part is for the
interference dimensions. The minus signs are a place holder that means ``no
signal''. The dimensions on the left $(\frac 12, 0, 0)$ are $(d_{1,1},
d_{1,2}, d_{1,3})$, the DoF's that receiver 1 needs. These entries have been
removed from the right part (replaced with minus signs). Counting the total
dimensions by taking the maximum of all the DoF on each column, treating minus
as 0, we have
\begin{equation}
  \frac 12 + 0 + 0 + \frac 12  + 0 + 0 = 1,
\end{equation}
which is acceptable. Similar verification can be performed for receiver 2 and
3 as well. As a result, the point as in \eqref{eq.half} is achievable.
Formally, we state without proof the following.
\begin{theorem} [Permuted Alignment] \label{th.perm}
For a $K$-transmitter $J$-receiver constant-coefficient single-antenna
wireless X network $\bH\in \setR^{J\times K}$, the DoF region $\cD$ satisfies
$\cD\supset \cDin_0$ a.e., where $\cDin_0$ is a set of matrices $[ d_{j,k}
]_{j=1,k=1}^{J,K} $ such that
\begin{enumerate}
\item All entries of it are non-negative;
\item There exists $K$ permutations of $J$ objects $\{\sigma_k(\cdot)|k\in\cK \}$ such
that $\forall 1\le \hj\le J$, the following inequality holds:
\begin{equation} \label{eq.perm}
  \sum_{\tk=1}^K d_{\hj,\tk} +
  \sum_{\tj\in \cJ} \max\cI_{\tj} \le 1,
\end{equation}
where
\(
  \cI_{\tj}\bydef \{ d_{j,k}|k\in \cK,
  j\in \cJ, j\ne \hj, \sigma_k(j)=\tj \}
\).
\end{enumerate}
\end{theorem}
It should be obvious that if we choose the permutations to be the identity
mapping (no permutation), then the result in \thmref{th.main} is recovered.
For the purpose of comparison, we will term the alignment scheme in
\secref{sec.dof} the \emph{natural alignment}.

\subsection{Staggered alignment}

In both the natural alignment and the permuted alignment, any message from any
single transmitter is aligned with one and only one message from another
transmitter. However, this can be generalized. It is possible to align two
users' messages so that one message from the first user is aligned with
multiple messages from the other user.

\begin{figure}
\centerline{\includegraphics[width=.55\linewidth]{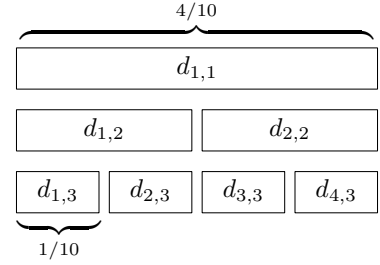}}
\caption{Staggered alignment}\label{fig.stag}
\end{figure}

Staggered alignment can achieve DoF point that are not achievable using the
natural or permuted alignments. Consider a $3\times 4$ X network. The point
$[d_{j,k}]$ as follows
\begin{equation}\label{eq.stag}
  [d_{j,k}]^T = \frac 1{10}\begin{bmatrix}
    4 & 0 & 0 & 0 \\
    2 & 2& 0 & 0\\
    1 & 1 & 1 & 1
  \end{bmatrix}
\end{equation}
is in the DoF region. This can be established using a staggered alignment
scheme as shown in \figref{fig.stag}.
Using permuted alignment without message staggering, a DoF point that is
proportional to the matrix in \eqref{eq.stag} will have a coefficient $
1/{11}$ instead of $1/{10}$ in front.

\section{Discussion} \label{sec.dis}

\subsection{Some special cases} An outer bound for the wireless X channel has
been derived in \cite{jash08}. It states that $\forall j\in \cJ$, $\forall
k\in \cK$:
\begin{equation}
\sum_{\tk=1}^K d_{j,\tk} + \sum_{\tj=1}^J d_{\tj, k} - d_{j,k} \le 1.
\end{equation}
This result can be written in an alternative form as
\begin{equation}\label{eq.alter}
\sum_{\tk=1}^K d_{j,\tk}+ \max_{k} \sum_{\tj\in \cJ, \tj\ne j}^J d_{\tj, k} \le 1,
\quad \forall j\in \cJ.
\end{equation}

\subsubsection{$K\times 2$ X channel} Comparing \eqref{eq.main} and
\eqref{eq.alter}, it can be seen that the inner bound does not meet the outer
bound in general. However, there are some special cases where they do meet.
One such case is when $J=2$. In this case, both bounds are given by
\begin{align}
  \sum_{\tk=1}^K d_{1,\tk}+ \max_{k} d_{2, k}  & \le 1,\label{eq.first}\\
  \sum_{\tk=1}^K d_{2,\tk}+ \max_{k} d_{1, k}  & \le 1. \label{eq.second}
\end{align}
We summarize the result in the following.
\begin{theorem}[DoF Region of $K\times 2$ X Network]
The DoF region of the $K\times J$ wireless X network when $J=2$ is a.e. the
set of $[d_{j,k}]_{j=1,k=1}^{2,K}$ that have non-negative entries and satisfy
both \eqref{eq.first} and \eqref{eq.second}.
\end{theorem}

\subsubsection{Some boundary points on the general DoF region} Another case
where the two bounds \eqref{eq.main} and \eqref{eq.alter} meet is when
$d_{j,k}=d_{j,\hk}$, for all $j\in \cJ$ and for all $k, \hk\in \cK$. We have:
\begin{theorem}[Some Boundary Points]
The DoF region of the $K\times J$ wireless X network a.e.\ has the following
points on the boundary: $[d_{j,k}]_{j=1,k=1}^{J,K}$ such that
\begin{enumerate}
\item [i)] all entries are non-negative;
\item [ii)] $d_{j,k}=d_{j,\hk}$, for all $j\in \cJ$ and for all $k, \hk\in
\cK$;
\item [iii)] $(K-1)\max_{j\in \cJ} d_{j,1}+\sum_{j\in \cJ} d_{j,1}=1$.
\end{enumerate}
This is true for Lebesgue almost everywhere $\bH\in \setR^{J\times K}$.
\end{theorem}
If we set all $d_{j,k}=1/(J+K-1)$, then we recover the total DoF of
$d^\text{(total)}={JK}/{(J+K-1)}$ of \cite{caja09,mgmk09}.

\subsection{Extensions} It is possible to extend the result in the paper to
cases where each transmitter emits an arbitrary number of messages, and each
receiver may request an arbitrary subset of the messages emitted by all the
transmitted. This can be termed the wireless X network with multicast, or
wireless X network with general message demands. For the case where each
transmitter emits only one message, and the channel varies with time, the DoF
region for time-varying channel has been obtained in \cite{krwy11i}. The same
DoF region, but for a constant coefficient channel, can be derived using the
technique developed in this paper. As a special case of that, the DoF region
result of $K$-user interference channel with single antennas \cite{wusv11} can
also be recovered.

\section{Conclusions}\label{sec.conc}

We have derived some achievability results for the wireless X network with
single antennas. Each message is split into multiple streams, and
achievability is established using real interference alignment of the streams.
The streams emitted by a single transmitter can be ``shuffled'' to determine
the alignment position with respect to streams from other transmitters. Such
rearrangement allow for higher DoF in some cases. It is not known whether the
presented schemes are sufficient to achieve all points in the DoF region.
However, we showed that when the number of receivers is equal to two, then the
achieved region is actually the DoF region. We also showed that certain
boundary points in the general DoF region can be achieved using the proposed
schemes. Closing the possible gap would be a meaningful objective. It would
also be interesting to investigate the multiple antenna cases. Quantifying the
DoF region of the general wireless X network with arbitrary number antennas at
each node, and with general message demands, would be a useful result.

\emph{Acknowledgement:} The work in this paper was supported in part by the
NSF grant No.~1128477.

\bibliographystyle{IEEE-unsorted}
\bibliography{refs}

\end{document}